\documentclass[12pt]{article}
\usepackage{fullpage}
\usepackage{url}
\usepackage{xspace} 
\usepackage{alltt}
\usepackage[T1]{fontenc}
\usepackage{array} 
\newcolumntype{L}[1]{>{\raggedright\let\newline\\\arraybackslash\hspace{0pt}}m{#1}}
\newcolumntype{C}[1]{>{\centering\let\newline\\\arraybackslash\hspace{0pt}}m{#1}}
\newcolumntype{R}[1]{>{\raggedleft\let\newline\\\arraybackslash\hspace{0pt}}m{#1}}

\renewcommand\b[1]{{\bf #1}} 
\newcommand\m[1]{\mbox{$#1$}} 
\renewcommand{\O}[1]{\m{O(#1)}} 
\newenvironment{code}{\begin{alltt}\footnotesize}{\end{alltt}}
\renewcommand\c[1]{\mbox{\tt\footnotesize #1}} 
\newcommand\p[1]{\mbox{\small\m{\it #1}}}
\newcommand{\IF}{<\hspace{-.0ex}{\normalsize-}} 

\newcommand{\notes}[1]{} 
\newcommand{\myparag}[1]{\subsection{#1}}
\newcommand{\descheading}[1]{\b{\hspace{2ex}\small #1.}}
\newcommand{\Fsp}[1]{\mbox{\hspace{#1ex}}} 
\newcommand{\Fsptwo}{\mbox{\hspace{2ex}}} 

\begin{document}

\title{Logic programming applications:\\ What are the abstractions and
  implementations?\thanks{This work was supported in part by NSF under
    grants CCF-0964196, 
    CCF-1248184, CCF-1414078, 
    and IIS-1447549; 
    and ONR under grants N00014-15-1-2208.} 
}

\date{December 24, 2017} 
\author{Yanhong A. Liu\\{\small \Fsp{1} Computer Science Department, Stony
    Brook University}\\{\small liu@cs.stonybrook.edu}}
\maketitle

\begin{abstract}

  This article presents an overview of applications of logic programming,
  classifying them based on the abstractions 
  and implementations of logic languages that support the applications.
  The three key abstractions are join, recursion, and constraint\notes{,
  capturing bounded, cyclic, and general computations, respectively}.
  Their essential implementations are for-loops, fixed points, and
  backtracking, respectively.
  The corresponding kinds of applications are database queries, inductive
  analysis,
  and combinatorial search, respectively.
  We also discuss language extensions and programming paradigms, summarize
  example application problems by application areas, and touch on example
  systems that support variants of the abstractions with different
  implementations.
\end{abstract}

\section{Introduction}

Common reasoning with logic is the root of logic programming, which allows
logic rules and facts to be expressed formally and used precisely for
inference, querying, and analysis in general.
Logic formalisms, or languages, allow complex application problems to be
expressed declaratively with high-level abstractions and allow desired
solutions to be found automatically with potentially efficient low-level
implementations.\notes{solvers, engines}

The biggest challenge in logic programming has been the need for efficient
implementations.  Much progress has been made,
with efficient implementations in some cases beating manually written
low-level code.  However, inadequate performance in many cases has led to
the introduction of non-declarative features in logic languages and
resulted in\notes{forced by me, promoted by Kifer, prompted by Scott} the
writing of\notes{many, changed from ``some'' by David Warren, then deleted
  by Kifer} obscure logic programs.

Despite the challenges, the most exciting aspect of logic programming is
its vast areas of applications.  They range from database queries to
program analysis, from text processing to decision making\notes{planning},
from security to knowledge engineering, and more.
These vast, complex, and interrelated areas make it challenging but 
necessary to provide a deeper understanding of the various kinds of
applications in order to help advance the state of the art of logic
programming and realize its benefits.

This article presents an overview of applications of logic programming
based on a study of the abstractions and implementations of logic
languages.  The rationale is that abstractions and implementations are the
enabling technologies of the applications.
The abstractions are essential for determining what kinds of application
problems can be expressed and how they can be expressed, for ease of
understanding, reuse, and maintenance.  The underlying implementations are
essential for high-level declarative languages to be sufficiently efficient
for 
substantial applications.

%
%
%
We discuss the following essential abstractions, where data abstractions
are for expressing the data, and control abstractions are for expressing
computations over the data:
\begin{enumerate}

\item data abstractions: objects and relationships\notes{, captured by
    primitive and structured values and predicates};

\item control abstractions: (1) join, (2) recursion, and (3) constraint,
  which capture bounded, cyclic, and general computations, respectively.

\end{enumerate}
In logic languages, the data abstractions as objects and relationships are
essential for all three control abstractions.

The essential techniques for implementing the three control abstractions
listed are (1) for-loops, (2) fixed points, and (3) backtracking,
respectively.  The corresponding kinds of applications are
\begin{description}

\item[~] (1)\, database-style queries, e.g., for ontology management,
  business intelligence, and access control;

\item[~] (2)\, inductive analysis, e.g., for text processing, program
  analysis, network traversal, and trust management;

\item[~] (3)\, combinatorial search, e.g., for decision
  making\notes{planning}, resource allocation, games and puzzles, and
  administrative policy analysis.

\end{description}
We categorize application problems using these three control abstractions
because they capture conceptually different kinds of problems, with
inherently different implementation techniques, and at the same time
correspond to very different classes of applications.

Note that the same application domain may use different 
abstractions and implementations for different 
problems.
For example, enterprise software may use all three of traditional database
queries, inductive analysis, and combinatorial search, for business
intelligence and decision making\notes{planning}; and security policy
analysis and enforcement may use database-style queries for access control,
inductive analysis for trust management, and combinatorial search for
administrative policy analysis.

We also discuss additional extensions, especially
regular-expression paths for higher-level queries
and updates for modeling actions; 
additional applications; and abstractions used in main programming
paradigms.
We also touch on several well-known systems while discussing the
applications.

There is a large body of prior work, including surveys of logic programming
in general and applications in particular, as discussed in
Section~\ref{sec-relate}.  This article distinguishes itself from past work
by analyzing classes of applications based on the language abstractions and
implementations used.

The rest of the article is organized as follows.  Section~\ref{sec-lang}
presents essential abstractions in logic languages.
Sections~\ref{sec-join}, \ref{sec-recursion}, and~\ref{sec-constraint}
describe abstractions, implementations, and applications centered around
join, recursion, and constraint. 
Section~\ref{sec-ext} discusses additional language extensions,
applications, and programming paradigms.  Section~\ref{sec-relate}
discusses related literature and \notes{concludes with }future directions.

\section{Logic language abstractions}
\label{sec-lang}
Logic languages provide very high-level data and control abstractions,
using mostly very simple language constructs.
We describe these abstractions and their meanings intuitively.

\myparag{Data abstractions}
%
All data in logic languages are abstracted, essentially, as objects and
relationships.
\begin{description}

\item \descheading{Objects}
  Objects are primitive\notes{ constant} values, such as numbers and
  strings, 
  or structured values whose components are\notes{ recursively} objects.

  Examples of primitive values are integer number \c{3} and string
  \c{'Amy'}.  We enclose a string value in single quotes; if a string
  starts with a lower-case letter, such as \c{'amy'}, the quotes can be
  omitted, as has been conventional in logic languages.

  Examples of structured values are \c{succ(3)}, \c{father(amy)},
  and\linebreak \c{cert('Amy',birth('2000-02-28','Rome'))}, denoting the
  successor integer of \c{3}, the father of \c{amy}, and the certificate
  that \c{'Amy'} was born on \c{'2000-02-28'} in \c{'Rome'}, respectively.

  The names of structures, such as \c{succ}, \c{father}, \c{cert}, and
  \c{birth} above, are called function symbols.
  They correspond to object constructors in object-oriented languages.


\item \descheading{Relationships}
  %
  Relationships are \notes{asserted }predicates, or properties, that hold
  among objects.  In particular, \c{\p{p}(\p{o_1},...,\p{o_k})}, i.e.,
  predicate \p{p} over objects \c{\p{o_1},...,\p{o_k}} being true, is
  equivalent to \c{(\p{o_1},...,\p{o_k}) in \p{p}}, i.e., tuple
  \c{(\p{o_1},...,\p{o_k})} belonging to relation \p{p}---a table that
  holds the set of tuples of objects over which \p{p} is true.

  Examples of relationships are \c{male(bob)}, \c{is\_parent(bob,amy)},
  and\linebreak \c{issue(mario,'Amy',birth('2000-02-28','Rome'))}, denoting
  that \c{bob} is male, \c{bob} is a parent of \c{amy},
  and \c{mario} issued a certificate that \c{'Amy'} was born on
  \c{'2000-02-28'} in \c{'Rome'}, respectively.

  Structured values can be easily captured using relationships, but not
  vice versa.  For example, \c{f} being the structured value \c{father(c)}
  can be captured using relationship \c{is\_father(f,c)},
  but relationship \c{is\_parent(p,c)} cannot simply be captured as \c{p}
  being the structured value \c{parent(c)} when
  \c{c} has two parents.
\end{description}
%
Such high-level data abstraction allows real-world objects or their
lower-level representations, from bits and characters to lists to sets, to
be captured easily without low-level implementation details.  For example,
\begin{itemize}
\item bits and characters are special cases of integers and strings,
  respectively.
\item lists are a special case of linearly nested structured values, and
\item sets are a special case of relations consisting of tuples of one
  component.
\end{itemize}
Objects and relationships can be implemented 
using well-known data structures, including linked list, array, hash table,
B-tree, and trie,
usually taking \O{1} or \O{\log n} time to access an object, where \m{n} is
the size of the data.

\myparag{Control abstractions}
\label{sec-control}
%
%
%
%
Control in logic languages is abstracted at a high level, as logical
inference or logic queries over asserted relationships among objects:
\begin{itemize}

\item asserted relationships can be connected by logical connectives:
  conjunction (read ``and''), disjunction (read ``or''), negation (read
  ``not''), implication (read ``then''), backward implication (read
  ``if''), and equivalence (read ``if and only if'');

\item variables can be used in place of objects and be quantified over with
  universal quantifier 
  (read ``all'') and existential quantifier 
  (read ``some''); and

\item one can either infer all relationships that hold or query about
  certain relationships, among all objects or among certain objects.

\end{itemize}

Rules and facts are the most commonly supported forms in existing logic
languages:
\begin{description}

\item \descheading{Rules}
  A rule is of the following form, where \p{assertion_0} is called the
  conclusion, and other assertions are called the hypotheses.  Each
  assertion is a predicate over certain objects, where variables may be
  used in place of objects.  Intuitively, left arrow (\c{\IF})
  indicates backward implication, comma (\c{,}) denotes conjunction, and all
  variables in a rule are implicitly universally quantified, i.e., the rule
  holds for all values of the variables.
\begin{code}
    \p{assertion\sb{0}} \IF \p{assertion\sb{1}}, ..., \p{assertion\sb{k}}.
\end{code}

  For example, 
  the second rule below says:
  \c{X} is a grandfather of \c{Y} if \c{X} is the father of \c{Z} and \c{Z}
  is a parent of \c{Y}, and this holds for all values of variables \c{X},
  \c{Y}, and \c{Z};
  the other rules can be read similarly.
  %
  Following logic language conventions, names starting with an upper-case
  letter are variables.
\begin{code}
    is_parent(X,Y) \IF is_father(X,Y).
    is_grandfather(X,Y) \IF is_father(X,Z), is_parent(Z,Y).
    is_ancestor(X,Y) \IF is_parent(X,Z), is_ancestor(Z,Y).
    is_positive(succ(N)) \IF is_positive(N).
\end{code}
  The second rule is 
  a join query---its two hypotheses have a shared variable, and it concludes
  a new predicate.

  The third and fourth rules are recursive---the predicate in the
  conclusion depends on itself in a hypothesis, or in general possibly
  indirectly through another predicate. 

  Note that disjunction of a set of hypotheses can be expressed using a set
  of rules with the same conclusion.

\item \descheading{Facts}
  A fact is a rule that has no hypotheses and is denoted simply as
  \c{assertion0.}
  For example, \c{is\_father(bob,amy).} says that \c{bob} is the father of
  \c{amy}, and \c{is\_positive(1).} says that \c{1} is positive.
\end{description}
The meaning of a set of rules and facts is the least set of facts
that contains all the given facts and all the facts that can be inferred,
directly or indirectly, using the rules.  This set can be computed by
starting with the given facts and repeatedly applying the rules to conclude
new facts---i.e.,
matching hypotheses of rules against facts, instantiating variables in
rules with values in matched facts, and adding instantiated conclusions of
rules as new facts.
\notes{
Formally, let $\theta$ be a substitution that maps variables to
constants, and $\theta(Rules)$ be the set of implications obtained by
applying $\theta$ to variables in $Rules$.  Then the meaning of a set
$Rules$ of rules and a set $Facts$ of facts can be defined as the
least fixed point of the following function $F$ that includes $Facts$:
\begin{equation}
\label{eq:F}
  F(S) = S\cup
  \{ fact~|~\exists\theta~(fact\mbox{\IF} fact_1,..., fact_h)\in\theta(Rules),
  fact_1\!\in S, ..., fact_h\!\in S \}
\end{equation}
}
However,
\begin{itemize}

\item repeated application of rules might not terminate if function symbols
  are used in the rules, because 
  facts about infinitely many new objects may be concluded, e.g., the
  fourth example rule above may infer \c{is\_positive(succ(1))},
  \c{is\_positive(succ(succ(1)))}, and so on.

\item when only certain relationships about certain objects are queried,
  application of rules may stop as soon as the query can be answered, e.g.,
  if only \c{is\_positive(succ(1))} is queried, application of rules can
  stop after one use of the given rule and the given fact.

\end{itemize}
Rules that do not contain function symbols 
are called Datalog rules.  For example, the first three example rules given
earlier in this section are Datalog rules.

General logic forms have also been increasingly supported, typically by
extending the rule form above:
\begin{description}
\item \descheading{Negation in the hypotheses}
  A hypothesis in a rule may be prefixed with \c{not}, denoting negation of
  the asserted relationship.

  For example, the following rule says: for all values of \c{X} and \c{Y},
  \c{X} is the mother of \c{Y} if \c{X} is a parent of \c{Y} and \c{X} is
  not male.
\begin{code}
    is_mother(X,Y) \IF is_parent(X,Y), not male(X).
\end{code}  

  Difficulties arise when negation is used with recursion. For example,
  what can be inferred from the following rule?  Is \c{good(zak)} true or
  false?
\begin{code}
    good(zak) \IF not good(zak).
\end{code}

\item  \descheading{More general forms}
  More general forms include disjunction and negation in the conclusion
  and, most generally, quantifiers \c{all} and \c{some} in any scope, not
  only the outermost scope.
  For example, the first rule below says: \c{X} is male or female if \c{X}
  is a person.  The second rule says: \c{X} is not a winning position if,
  for all \c{Y}, there is no move from \c{X} to \c{Y} or else \c{Y} is a
  winning position.
\begin{code}
    male(X) or female(X) \IF person(X).
    not win(X) \IF all Y: not move(X,Y) or win(Y).
\end{code}
\notes{
  Disjunction in the conclusion.  This stands out as missing because
  disjunction of hypotheses can be captured by multiple rules the same
  conclusion and each with one of the disjuncts as the only hypothesis, and
  conjunction in the conclusion can be captured by multiple rules with the
  same hypotheses and each with one of the conjuncts as the conclusion.

  For example, the following rule says: \c{X} is male or female if
  \c{X} is a person.
\begin{code}
    male(X) or female(X) \IF person(X).
\end{code}
  In most ASP systems,

  Negation in the conclusion.
  Not directly in ASP, indirectly by constraints, and in FO/ID.


  General quantifications.  in any scope, not just the outermost scope.
  This is most general.  

  Conjunction, negation, and one quantifier in any scope suffice to
  capture all of predicate logic.

  For example, \c{\m{\forall} X: male(X), female(X)}

  not in ASP, but in FO/ID.
}

\end{description}
The meaning of recursive rules with negation is not universally agreed
upon.  The two dominant semantics are well-founded semantics
(WFS)~\cite{van+91well,van93alt} and stable model semantics
(SMS)~\cite{GelLif88}.
Both WFS and SMS use the closed-world assumption, 
i.e., they assume that what cannot be inferred to be true from the given
facts and rules, is false.
\begin{itemize}

\item WFS gives a single 3-valued model, with the additional truth value
  \c{undefined} besides \c{true} and \c{false}.

\item SMS gives zero or more 2-valued models, using only \c{true} and
  \c{false}.

\end{itemize}
%
Other formalisms and semantics include partial stable models, also called
stationary models~\cite{prz94well};
first-order logic with inductive and fixed-point definitions, called FO(ID)
and FO(FD)~\cite{denecker2008logic,hou2010fo}; and the newly proposed
founded semantics and constraint semantics~\cite{LiuSto18Founded-LFCS}.
The first two are both aimed at unifying WFS and SMS.
The last unifies and cleanly relates WFS, SMS, and other major semantics by
allowing the assumptions about the predicates and rules to be specified
explicitly.

For practical applications, logic languages often also support
predefined\notes{built-in} relationships among objects, including equality,
inequality, and general comparisons.  Cardinality and other aggregates over
relationships are often also supported.
%

\myparag{Combinations of control abstractions}
There are many possible combinations of the language constructs.  We focus
on the following three combinations of constructs as essential control
abstractions.  We identify them by join, recursion, and constraint.  They
capture bounded, cyclic, and general computations, respectively.
\begin{description}

\item[] (1)\,\, \b{Join}---with join queries, no recursive rules, and
  restricted negation and other constructs; the restriction is that,
  for each rule, 
  each variable in the conclusion must also appear in a hypothesis that is
  a predicate over arguments.
  Implementing this requires that common objects for the shared variables
  be found for the two hypotheses of a join query to be true at the same
  time; the number of objects considered are bounded, by the predicates in
  the hypotheses, following a bounded number of dependencies.

\item[] (2)\,\, \b{Recursion}---with join queries, recursive rules, and
  restricted negation and other constructs; the restriction is as for join
  above plus that a predicate in the conclusion of a rule 
  does not depend on
  the negation of the predicate itself in a hypothesis.
  Implementing this requires repeatedly applying the recursive rules
  following cyclic dependencies, potentially an unbounded number of times
  if new objects are in some conclusions.

\item[] (3)\, \b{Constraint}---with join queries, recursive rules, and
  unrestricted negation and other constructs; unrestricted negation and
  other constructs can be viewed as constraints to be satisfied.
  Implementing this could require, in general, trying different
  combinations of variable values, as in general constraint solving.

\end{description}
Table~\ref{tab-abs} summarizes these three essential control abstractions
and the corresponding kinds of computations and applications.
\begin{table}[htbp]
  \centering
\begin{tabular}[t]{@{}l@{~}|l@{~}||l@{~}|l@{~}|l@{~}||l@{~}||@{~}l@{}}\hline
     & Essential 
     & \begin{tabular}[c]{@{}l@{}} Has join\\ queries \end{tabular}
     & \begin{tabular}[c]{@{}l@{}} Has rec.\\ rules \end{tabular}
     & \begin{tabular}[c]{@{}l@{}} Has neg.\\ and others \end{tabular}
     & Computations & Application kinds\\
       \hline\hline                             
  (1)& Join       & yes & no  & restricted   & bounded & database-style queries\\\hline
  (2)& Recursion  & yes & yes & restricted   & cyclic  & inductive analysis\\\hline
  (3)& Constraint & yes & yes & unrestricted & general & combinatorial search\\\hline
\end{tabular}
\caption{Essential control abstractions of logic languages.}
\label{tab-abs}
\end{table}

\section{Join and database-style queries}
\label{sec-join}
%
%
%
%
%

Join queries are the most basic and most commonly used queries in relating
different objects.  They underlie essentially all nontrivial queries in
database applications and many other applications.

\myparag{Join queries}
%
A join query is a conjunction of two hypotheses that have shared variables,
concluding 
possible values of variables that satisfy both hypotheses.
A conjunction of two hypotheses that have no shared variables, i.e., a
Cartesian product, or a single hypothesis can be considered a trivial join
query.
A join query corresponds to a rule whose predicate in the conclusion is
different from predicates in the hypothesis, so the rule is not recursive.
A non-recursive rule with more than two hypotheses corresponds to multiple
join queries, as a nesting or chain of join queries starting with joining
any two hypotheses first.

For example, the first rule below, as seen before, is a join query.  So is
the second rule; it defines \c{sibling} over \c{X} and \c{Y} if \c{X} and
\c{Y} have a same parent.  The third rule defines
a chain of red, green, and blue links from \c{X} to \c{Y} through \c{U} and
\c{V};
it can be viewed as two join queries---join any two hypotheses first, and
then join the result with the third hypothesis.
\begin{code}
    is_grandfather(X,Y) \IF is_father(X,Z), is_parent(Z,Y).
    sibling(X,Y) \IF is_parent(Z,X), is_parent(Z,Y).
    chain(X,Y) \IF link(X,U,red), link(U,V,green), link(V,Y,blue).
\end{code}
%
%
%
In general, the asserted predicates can be about relationships among any
kinds of objects---whether people, things, events, or anything else, e.g.,
students, employees, patients, doctors, products, courses, hospitals,
flights, interviews, and hangouts; and the join queries can be among any
kinds of relationships---whether family, friend, owning, participating,
thinking, or any other relation in the real world or conceptual world.

Join queries expressed using rules correspond to set queries.  For example,
in a language that supports set comprehensions with tuple
patterns~\cite{RotLiu07Retrieval-PEPM,Liu+16IncOQ-PPDP} the
\c{is\_grandfather} query corresponds to
\begin{code}
    is_grandfather = \{(X,Y): (X,Z) in is_father, (Z,Y) in is_parent\}
\end{code}
Without recursion, join queries can be easily supported together with the
following extensions, with the restriction that,
for each rule, each variable in the conclusion must also appear in a
hypothesis that is a predicate over arguments, so the domain of the
variable is bounded by the predicate;
queries using these extensions can be arbitrarily nested:
\begin{itemize}

\item unrestricted negation, other connectives, and predefined relationships
  in additional conditions,

\item aggregates, such as count and max, about the relationships, and

\item general universal and existential quantifiers in any scope.

\end{itemize}
These subsume all constructs in the \c{select} statement for SQL queries.
Essentially, join queries, with no recursion, relate objects in different
relationships within a bounded number of steps.

\myparag{Implementation of join queries}
\label{sec-join-impl}
A join query can be implemented straightforwardly using nested for-loops
and if-statements, where shared variables in different hypotheses
correspond to equality tests between the corresponding variables.
For example, the \c{is\_grandfather} query earlier in this section can be
implemented as
\begin{code}
    is_grandfather = \{\}
    for (X,Z1) in is_father:         -- time factor: number of is_father pairs
      for (Z2,Y) in is_parent:       -- time factor: number of is_parent pairs
        if Z1 == Z2: 
          is_grandfather.add(X,Y)
\end{code}
In a language that supports set comprehensions, such as Python, the above
implementation can be expressed as
\begin{code}
    is_grandfather = \{(X,Y) for (X,Z1) in is_father for (Z2,Y) in is_parent if Z1 == Z2\}
\end{code}

For efficient implementations, several key implementation and optimization
techniques are needed, described below; additional optimizations are also
needed, e.g., for handling streaming data or distributed data.
\begin{description}

\item \descheading{Indexing}
  This creates an index for fast lookup based on values of the indexed
  arguments of a relation; the index is on the shared arguments of the two
  hypotheses.  For example, for any fact \c{is\_father(X,Z)}, to find the
  matching \c{is\_parent(Z,Y)}, an index called, say,
  \c{children\{Z\}}---mapping the value of \c{Z}, the first argument of
  \c{is\_parent}, to the set of corresponding values of second argument of
  \c{is\_parent}---significantly speeds up the lookup,
  improving the time factor for the inner loop to the number of children of
  \c{Z}:
\begin{code}
    is_grandfather = \{(X,Y) for (X,Z) in is_father for Y in children\{Z\}\}
\end{code}

\item \descheading{Join ordering}
  This optimizes the order of joins when there are multiple joins, e.g., in
  a rule with more than two hypotheses.
  For example, for the rule for \c{chain}, starting by joining the first
  and third hypotheses is never more efficient than starting by joining
  either of these hypotheses with the second hypothesis, because the former
  yields all pairs of red and blue links, even if there are no green links
  in the middle.
  %

\item \descheading{Tabling}
  This stores the result of common sub-joins so they are not repeatedly
  computed.  Common sub-joins may arise when there are nested or chained
  join queries.
  For example, for the rule for \c{chain} earlier in this section, consider
  joining the first two
  hypotheses 
  first: if there are many red and green link pairs from a value of \c{X}
  to a value of \c{V},
  then storing the result of this sub-join avoids recomputing it when
  joining with blue links to find each target \c{Y}.

\item \descheading{Demand-driven computation}
  This computes only those parts of relationships that affect a particular
  query.  For example, a query may only check whether
  \c{is\_father(dan,bob)} holds, or find all values of \c{X} for
  \c{is\_father(dan, X)}, or find all \c{is\_father} pairs, as opposed to
  finding all relationships that can be inferred.

\end{description}
Basic ideas for implementing the extensions negation, aggregates, etc.\ are
as follows, where nested queries using these extensions are computed
following their order of dependencies:
\begin{itemize}

\item negation, etc. in additional conditions: test them after the
  variables in them become bound by the joins.

\item aggregates: apply the aggregate operation while collecting the query
  result of its argument.

\item quantifiers: transform them into for-loops, or into aggregates, e.g.,
  an existential quantification is equivalent to a count being positive.

\end{itemize}
Efficient implementation techniques for join queries and extensions have
been studied in a large literature, e.g.,~\cite{ioa1996query}.
Some methods also provide precise complexity guarantees,
e.g.,~\cite{wil02jcss,Liu+16IncOQ-PPDP}.

\myparag{Applications of join queries}
Join queries are fundamental in querying complex relationships among
objects.  They are the core of 
database applications~\cite{KifBL06},
from enterprise management to ontology management, from accounting systems
to airline reservation systems, and from electronic health records
management to social media management.
Database and logic programming are so closely related that one of the most
important computer science bibliographies is called DBLP, and it was named
after Database and Logic Programming~\cite{ley2002dblp}.
Join queries also underlie applications that do not fit in traditional
database applications, such as complex access control policy
frameworks~\cite{ansi04role}.

We describe three example applications below, in the domains of ontology
management, enterprise management, and security policy frameworks.
They all heavily rely on the use of join queries and optimizations,
especially indexing.  We give specific examples of facts, rules, and
indexing for the first application.
\begin{description}

\item \descheading{Ontology management---Coherent definition framework
    (CDF)}
  CDF is a system for ontology management that has been used in numerous
  commercial projects~\cite{Gomes2010mknf}, for organizing information
  about, e.g., aircraft parts, medical supplies, commercial processes, and
  materials.
  It was originally developed\notes{ as a proprietary tool} by XSB, Inc.
  Significant portions have been\notes{ made open source} released in
  the XSB packages~\cite{xsb14}.

  The data in CDF are classes and objects.  For example, XSB, Inc.\ has a
  part taxonomy, combining UNSPSC (United Nations Standard Products and
  Services Code)
  and Federal INC (Item Name Code) 
  taxonomies,
  with a total of over 87,000\notes{from David Warren on his laptop May 28,
    2015; was 70,000 in his previous email; was 60 in paper below}
  classes\notes{categories in paper and by warren, but warren confirmed
    they are classes} of parts.
  The main relationships are variants of \c{isa}, \c{hasAttr}, and
  \c{allAttr}.
  Joins\notes{ of these relations} are used extensively to answer queries
  about closely related classes, objects, and attributes.  Indexing and
  tabling are heavily used for efficiency.  Appropriate join order and
  demand-driven computation are also important.

  An example fact is as follows, indicating that specification
  \c{'A-A-1035'} in ontology \c{specs} has attribute \c{'MATERIAL'} whose
  value is\notes{ a specific aluminum alloy} \c{'ALUMINUM ALLOY UNS
    A91035'} in 
  \c{material\_taxonomy}.
  Terms \c{cid(Identifier, Namespace)} represent primitive classes in CDF.
\begin{code}
    hasAttr_MATERIAL(cid('A-A-1035',specs),
                     cid('ALUMINUM ALLOY UNS A91035',material_taxonomy)).
\end{code}
%
  An example rule is as follows, meaning that a part
  \c{PartNode} has attribute\linebreak \c{'PART-PROCESS-MATERIAL'} whose value is
  process-material pair \c{(Process,Material)} in\linebreak \c{'ODE
    Ontology'} if \c{PartNode} has attribute \c{'PROCESS'} whose value is
  \c{Process}, and \c{Process} has attribute \c{'PROCESS-MATERIAL'} whose
  value is \c{Material}.
\begin{code}
    hasAttr_PART-PROCESS-MATERIAL(PartNode,
            cid('process-material'(Process,Material),'ODE Ontology')) \IF
      hasAttr_PROCESS(PartNode, Process),
      hasAttr_PROCESS-MATERIAL(Process, Material).
\end{code}
%
An example of indexing is for \c{hasAttr\_ATTR}, for any \c{ATTR}, shown
below, in XSB notation, meaning: use as index all symbols of the first
argument if it is\notes{ (essentially) ground} bound, or else do so for the
second argument.
\begin{code}
    [*(1), *(2)]    
\end{code}
  XSB, Inc.\ has five major ontologies represented in CDF, for parts, 
  materials, etc., with a total of over one million facts and five
  meta\notes{ ontology-specific} rules. 
  The rules are represented using a Description Logic form---an ontology
  representation language.  The example rule above is an instance of 
  such a rule when interpreted.  The indexing used
  supports different appropriate indices for different join queries.

  CDF is used in XSB, Inc.'s ontology-directed classifier (ODC) and
  extractor (ODE)~\cite{swift2012xsb}.
  ODC uses a modified Bayes classifier to classify item descriptions.
  For example, it is\notes{checked with Warren May 28,2015} used quarterly
  by the U.S.\ Department of Defense to classify over 80 million part
  descriptions.  ODE extracts attribute-value pairs from classified
  descriptions to build structured knowledge about items\notes{ and their
    attributes}.  ODC uses aggregates extensively, and ODE uses string
  pattern rules.
  

\item \descheading{Enterprise management---Business intelligence (BI)}
  BI is a central component of enterprise software.
  It tracks the performance of an enterprise over time by storing and
  analyzing historical information recorded through online transaction
  processing (OLTP), and is then used to help plan\notes{decide} future
  actions of the enterprise.  LogicBlox simplifies the hairball of
  enterprise software technologies by using a Datalog-based
  language~\cite{green12logicblox,aref2015design}.

  All data are captured as logic relations.  This includes not only data as
  in conventional databases, e.g., sale items, price, and so on for a
  retail application, but also data not in conventional databases, e.g.,
  sale forms, display texts, and submit buttons in a user interface.  Joins
  are used for easily querying interrelated data, as well as for generating
  user interfaces.  Many extensions such as aggregates are also used.  For
  efficiency, exploiting the rich literature of automatic optimizations,
  especially join processing strategies and incremental
  maintenance, 
  is of paramount importance.


  Using the same Datalog-based language, LogicBlox supports not only BI but
  also OLTP and prescriptive and predictive analytics.  ``Today, the
  LogicBlox platform has matured to the point that it is being used daily
  in dozens of mission-critical applications in some of the largest
  enterprises in the world, whose aggregate revenues exceed
  \$300B''~\cite{aref2015design}.



  %

\item \descheading{Security policy frameworks---Core role-based access
    control (RBAC)}
  RBAC is a framework for controlling user access to resources based on
  roles.  It became an ANSI standard~\cite{ansi04role} 
  building on much research during the preceding decade and earlier,
  e.g.,~\cite{Landwehr:Heitmeyer:McLean:84,Ferraiolo+92,Gavrila:Barkley:98,ferraiolo01proposed}.
  %

  Core RBAC defines\notes{ core functions on} users, roles, objects,
  operations, permissions, sessions and a number of relations among these
  sets; the rest of RBAC adds a hierarchical relation over roles, in
  hierarchical RBAC, and restricts the number of roles of a user and of a
  session, in constrained RBAC.
  %
  Join queries are used for all main system functions, especially the
  \c{CheckAccess} function, review functions, and advanced review functions
  on the sets and relations.
  They are easily expressed using logic
  rules~\cite{barker+04effi,barker+06term}. 

  Efficient implementations rely on all main optimizations discussed,
  especially auxiliary maps for indexing and
  tabling~\cite{Liu+06ImplCRBAC-PEPM}.
  Although the queries are like relational database queries, existing
  database implementations would be too slow for functions like
  \c{CheckAccess}.
  Unexpectedly, uniform use of relations and join queries also led to a
  simplified specification, with unnecessary mappings removed, undesired
  omissions fixed, and constrained RBAC drastically
  simplified~\cite{LiuSto07RBAC-ONR}.

\end{description}

\section{Recursion and inductive analysis}
\label{sec-recursion}
%
%
%
%

Recursive rules 
are most basic and essential in relating objects that are an unknown number
of relationships apart.  They are especially important for problems that
may require performing 
the inference or queries %
a non-predetermined number of steps, depending on the data.

\myparag{Recursive rules and queries}
%
Given a set of rules, a predicate \p{p} depends on a predicate \p{q} if
\p{p} is in the conclusion of a rule, and either \p{q} is in a hypothesis
of the rule or some predicate \p{r} is in a hypothesis of the rule and
\p{r} depends on \p{q}.
A given set of rules is recursive if a predicate \p{p} in the conclusion of
a rule depends on \p{p} itself\notes{ in the set of rules}.


For example, the second rule below, as seen in Section~\ref{sec-control},
is recursive; the first rule is not recursive; the set of these two rules
is recursive, where the first rule is the base case, and the second rule is
the recursive case.
\begin{code}
    is_ancestor(X,Y) \IF is_parent(X,Y).
    is_ancestor(X,Y) \IF is_parent(X,Z), is_ancestor(Z,Y).
\end{code}
%
%
In general, recursively asserted relationships can be between objects of
any kind, e.g., relatives and friends that are an unknown number of
connections apart in social networks, %
direct and indirect prerequisites of courses in
universities, 
routing paths in computer networks, %
nesting of parts in products, supply chains in supply and demand networks,
transitive role hierarchy relation in RBAC, 
and repeated delegations in trust management systems. 

Recursive queries with restricted negation correspond to least fixed-point
computations.  For example, in a language that supports least fixed points,
the \c{is\_ancestor} query corresponds to the minimum \c{is\_ancestor} set
below, where, for any sets \c{S} and \c{T}, \c{S subset T} holds iff every
element of \c{S} is an element of \c{T}:
\begin{code}
    min is_ancestor: is_parent subset is_ancestor,
                     \{(X,Y): (X,Z) in is_parent, (Z,Y) in is_ancestor\} subset is_ancestor
\end{code}
With cyclic predicate dependencies, recursion allows the following restricted
extensions to be supported while still providing a unique semantics; there
is also the restriction that, 
for each rule, 
each variable in the conclusion must also appear in a hypothesis that is
a predicate over arguments,
as in extensions to join queries:
\begin{itemize}

\item stratified negation, where negation and recursion are
  separable\notes{separate into stratus}, i.e., there is no predicate that
  depends on the negation of itself\notes{too loose: negation in dependency
    cycles}, and

\item other connectives and predefined relationships in additional
  conditions, aggregates, and general quantifiers, as in extensions for
  join queries, when they do not affect the stratification.

\end{itemize}
Essentially, recursive rules capture an unbounded number of joins,
and allow inference and queries by repeatedly applying the rules.

\myparag{Implementation of recursive rules and queries}
Inference and queries using recursive rules can be implemented using
while-loops;
for-loops with predetermined number of iterations do not suffice, because
the number of iterations depends on the rules and facts.
Each iteration applies the rules in one step, so to speak, until no more
relevant facts can be concluded.
For example, the \c{is\_ancestor} query earlier in this section can be
implemented as
\begin{code}
    is_ancestor = is_parent
    while exists (X,Y): (X,Z) in is_parent, (Z,Y) in is_ancestor, (X,Y) not in is_ancestor:
      is_ancestor.add((X,Y))
\end{code}
Each iteration computes the existential quantification in the condition of
the while-loop, and picks any witness \c{(X,Y)} to add to the result set.
It can be extremely inefficient to recompute the condition in each
iteration after a new pair is added.

For efficient implementations, all techniques for joins are needed but are
also more critical and more complex.  In particular, to ensure termination,
\begin{itemize}
\item tabling is critical if relationships form cycles, and 
\item demand-driven computation is critical if new objects are created in
  the cycles.
\end{itemize}
For the \c{is\_ancestor} query, each iteration computes the following set,
which is a join, plus the last test to ensure that only a new fact is added:
\begin{code}
    \{(X,Y): (X,Z) in is_parent, (Z,Y) in is_ancestor, (X,Y) not in is_ancestor\}
\end{code}
Two general principles underlying the optimizations for efficient
implementations are:
\begin{enumerate}

\item incremental computation for expensive relational join operations,
  with respect to facts that are added in each iteration.

\item data structure design for the relations, for efficient retrievals and
  tests of relevant facts.

\end{enumerate}
For the restricted extensions, iterative computation follows the order of
dependencies determined by stratification; additional aggregates, etc.\
that do not affect the stratification can be handled as described in
Section~\ref{sec-join-impl} for computing the join in each iteration.

Efficient implementation techniques for recursive queries and extensions
have been studied extensively, e.g.,~\cite{AbiHulVia95}.  Some methods also
provide precise complexity guarantees,
e.g.,~\cite{mcallester99,Ganzinger:McAllester:01,LiuSto09Rules-TOPLAS,TekLiu11RuleQueryBeat-SIGMOD}.

\myparag{Applications of recursive rules and queries}
Recursive rules and queries can capture any complex reachability problem
in recursive structures, graphs, and hyper-graphs.
Examples are social network analysis based on all kinds of social graphs;
program analysis over many kinds of flow and dependence graphs about
program control and data values;
model checking over labeled transition systems and state machines;
\notes{network }routing in electronic data networks, telephone networks, or
transportation networks; and security policy analysis and enforcement over
trust or delegation relationships.

We describe three example applications below, in the domains of text and
natural language processing, program analysis, and distributed security
policy frameworks.  They all critically depend on the use of recursive
rules and efficient implementation techniques, especially tabling and
indexing.
%

\begin{description}

\item \descheading{Text processing---Super-tokenizer}
  Super-tokenizer is an infrastructure tool for text processing that has
  been used by XSB, Inc.'s ontology-directed classifier (ODC) and extractor
  (ODE) for complex commercial
  applications~\cite{swift2012xsb}.\notes{David ok-ed May 21, 2015}
  It was also developed originally at XSB, Inc.
  \notes{not true:and later made open source in XSB packages~\cite{xsb14}}

  %
  Super-tokenizer supports the declaration of complex rewriting rules for
  token lists.  For example, over 65,000\notes{from David Warren on his
    laptop May 28, 2015; in David's previous email: 100,000; in paper: tens
    of thousands} of these rules implement abbreviations and token
  corrections in ODC and complex pattern-matching rules in ODE for
  classification and extraction based on combined UNSPSC and Federal INC
  taxonomies at XSB, Inc.
  Recursion is used extensively in the super-tokenizer, for text parsing
  and processing.
  The implementation uses tabled grammars and trie-based indexing in
  fundamental ways.\notes{as well as negation to implement preferred
    rewritings.}
  %

  Super-tokenizer is just one particular application that relies on
  recursive rules for text processing and, more generally, language
  processing.  Indeed, the original application of Prolog, the first
  and main logic programming language, was natural language processing
  (NLP)~\cite{pereira2002prolog},
  and a more recent application in NLP helped the IBM Watson question
  answering system win the Jeopardy Man vs.\ Machine Challenge by defeating
  two former grand champions\notes{, Ken Jennings and Brad Rutter, on
    February 14-16,} in 2011~\cite{lally2011natural,lally2012question}.

\item \descheading{Program analysis---Pointer analysis}  
  Pointer analysis statically determines the set of objects that a pointer
  variable or expression in a program can refer to.  It is a fundamental
  program analysis with wide applications
  and has been studied extensively, e.g.,~\cite{hind01,sri2013alias}.  The
  studies especially include significantly simplified
  specifications using Datalog in more recent years,
  e.g.,~\cite{smara15pointer},
  and powerful systems such as bddbddb~\cite{whaley2005using} 
  and Doop~\cite{bravenboer2009strictly}, the latter built using
  LogicBlox~\cite{green12logicblox,aref2015design}.

  Different kinds of program constructs and analysis results relevant to
  pointers are relations.  Datalog rules capture the analysis directly as
  recursively defined relations.
  %
  For example, the well-known Andersen's pointer analysis for C programs
  defines a points-to relation based on four kinds of assignment
  statements~\cite{and94thesis}, leading directly to four Datalog
  rules~\cite{SahaRam05}.
  Efficient implementation critically depends on tabling, indexing, and
  demand-driven computation~\cite{SahaRam05,TekLiu11RuleQueryBeat-SIGMOD}.
  Such techniques were in fact followed by hand to arrive at the first
  ultra fast analysis~\cite{HeiTar01ultra,HeiTar01demand}.

  Indeed, efficient implementations can be generated from Datalog rules
  giving much better, more precise complexity
  guarantees~\cite{LiuSto09Rules-TOPLAS,TekLiu11RuleQueryBeat-SIGMOD} than
  the worst-case complexities, e.g., the well-known cubic time for
  Andersen's analysis.
  Such efficient implementation with complexity guarantees can be obtained
  for program analysis in general~\cite{mcallester99}.  Commercial tools
  for general program analysis based on Datalog have also been built, e.g.,
  by Semmle 
  based on CodeQuest~\cite{haj+2006codequest}.

\item \descheading{Security policy frameworks---Trust management (TM)}
  TM is a unified approach to specifying and enforcing security policies in
  distributed systems~\cite{BlaFeiLac96,GraSlo00,ruohomaa2005trust}.  It
  has become increasingly important as systems become increasingly
  interconnected, and logic-based languages have been used increasingly for
  expressing TM policies~\cite{bonatti2010datalog},
  e.g., SD3~\cite{Jim01}, RT~\cite{LiMitWin02}, Binder~\cite{DeT02},
  Cassandra~\cite{BecSew04}, and many extensions,
  e.g.,~\cite{becker2012foundations,sultana2013selective}.

  Certification, delegation, authorization, etc.\ among users, roles,
  permissions, etc.\ are relations.  Policy rules correspond directly to
  logic rules.  The relations can be transitively defined, yielding
  recursive rules.
  For example, one of the earliest TM frameworks, SPKI/SDSI~\cite{Ell+99},
  for which various sophisticated methods have been studied, corresponds
  directly to a few recursive rules~\cite{Hri+07SPKI-PPDP},
  and efficient implementations with necessary indexing and tabling were
  generated automatically.
  %

  TM studies have used many variants of Datalog with restricted
  constraints~\cite{LiMit03}, not unrestricted negation.
  A unified framework with efficient implementations is still lacking.
  %
  For example, 
  based on the requirements of the U.K.\ National Health Service, a formal
  electronic health records (EHR) policy was written, as 375 rules in
  Cassandra~\cite{Bec05ehr},
  heavily recursive.
  As the largest case study in the TM literature, its implementation was
  inefficient and incomplete---techniques like indexing were deemed needed
  but missing~\cite{Bec05thesis}.

\end{description}

\section{Constraint and combinatorial search}
\label{sec-constraint}
%
%
%
%
%

Constraints are the most general form of logic specifications, which easily
captures the most challenging problem-solving activities such as planning
and resource allocation\notes{scheduling}.

\myparag{Constraint satisfaction}
%
A constraint is, in general, a relationship among objects but especially
refers to cases when it can be satisfied with different choices of objects
and the right choice is not obvious.

For example, the 
rule below says that \c{X} is a winning position if there is a move from
\c{X} to \c{Y} and \c{Y} is not a winning position.
It states a relationship among objects, but its meaning is not obvious,
because the concluding predicates are recursively defined using a negation
of the predicate itself.
\begin{code}
    win(X) \IF move(X,Y), not win(Y).
\end{code}
%
%
%
%
%
%
In general, constraints can capture any real-world or conceptual-world
problems, e.g., rules for moves in any game---whether recreational,
educational, or otherwise; 
actions with conditions and effects for any planning activities;
participants and resource constraints in scheduling---whether for
university courses or manufacturer goods production or hospital surgeries;
real-world constraints in engineering design; as well as knowledge and
rules for puzzles and brain teasers.

Given constraints may have implications that are not completely explicit.
For example, the \c{win} rule implies not just the first constraint below,
but also the second, by negating the conclusion and hypotheses in the given
rule, following the closed-world assumption; the second constraint makes
the constraint about \c{not win} explicit:
\begin{code}
    win(X) if some Y: move(X,Y) and not win(Y)
    not win(X) if all Y: not move(X,Y) or win(Y)
\end{code}
Indeed, with general constraints, objects can be related in all ways using
all constructs together with join and recursion: unrestricted negation, other
connectives, predefined relationships, aggregates, and general quantifiers
in any scope.

%
However, due to negation in dependency cycles, the meaning of the rules and
constraints is not universally agreed on anymore.
\begin{itemize}

\item Well-founded semantics (WFS) gives a single, 3-valued model, where
  relationships that are true or false are intended to be supported from
  given facts, i.e., well-founded, and the remaining ones are
  \c{undefined}.

\item Stable model semantics (SMS) gives zero or more 2-valued models,
  where each model stays the same, i.e., is stable, when it is used to
  instantiate all the rules;
  in other words, applying the rules to each model yields the same model.

\end{itemize}
For example, for the \c{win} example,

\begin{itemize}

\item if there is only one move, \c{move(a,b)}, not forming a cycle, then
 
  WFS and SMS both give that \c{win(b)} is false and \c{win(a)} is true;


\item if there is only one move, \c{move(a,a)}, forming a self cycle, then

  WFS gives that \c{win(a)} is undefined, and

  SMS gives that there is no model;

\item if there are only two moves, \c{move(a,b)} and \c{move(b,a)}, forming
  a two-move cycle, then

  WFS gives that \c{win(a)} and \c{win(b)} are both undefined, and

  SMS gives two models: one with \c{win(a)} true and \c{win(b)} false, and
  one with the opposite results.

\end{itemize}
Despite the differences, WFS and SMS can be computed using some shared
techniques.
%

\myparag{Implementation of constraint satisfaction}
%
Constraint solving could in general use straightforward
generate-and-test---%
generate each possible combination of objects for solutions and test
whether they satisfy the constraints---but backtracking is generally used,
as it is much more efficient.
\begin{description}

\item \descheading{Backtracking}
  Backtracking incrementally builds variable assignments for the solutions,
  and abandons each partial assignment as soon as it determines that the
  partial assignment cannot\notes{ possibly} be completed to a satisfying
  solution, going back to try a different value for the last variable
  assigned; 
  this avoids trying all possible ways of completing those partial
  assignments or naively enumerating all complete assignments.
%
  
\end{description}
For example, the \c{win(X)} query can basically try a move at each next
choice of moves and backtrack to try a different move as soon as the
current move fails.  Expressed using recursive functions, this corresponds
basically to the following:
\begin{code}
    def win(X): return (some Y: move(X,Y) and not_win(Y))
    def not_win(X): return (all Y: not move(X,Y) or win(Y))
\end{code}
%
%
%
%
This backtracking answers the query correctly when the moves do not form a
cycle.  However, it might not terminate when the moves form a cycle, and
the implementation depends on the semantics used.
Both WFS and SMS can be computed by using and extending the basic
backtracking:
\begin{itemize}

\item WFS computation could track cycles, where executing a call requires
  recursively making the same call, and infer undefined for those %
  queries 
  that have no 
  execution paths to 
  infer the query result to be true or false.

\item SMS computation could generate possible partial or complete variable
  assignments, called grounding, and check them,
  possibly with the help of
  an external solver like Boolean satisfiability (SAT) solvers 
  or satisfiability modulo theories (SMT) solvers. 

\end{itemize}

For efficient implementations, techniques for join and recursion are
critical as before, especially tabling to avoid repeated states in the
search space.
Additionally, good heuristics for pruning the search space can make drastic
performance difference in computing SMS, e.g., as implemented in answer set
programming (ASP) solvers. 
\begin{description}


\item \descheading{Backjumping} 
  One particular optimization of backtracking in SMS computation is
  backjumping.  Backtracking always goes back one level in the search tree
  when all values for a variable have been tested.  Backjumping may go back
  more levels, 
  by realizing that a prefix of the partial assignment can
  lead to all values for the current variable to fail.  This helps prune
  the search space.


\end{description}
For extensions that include additional constraints, such as integer
constraints, as well as aggregates and quantifiers, an efficient solver
such as one that supports mixed integer programming (MIP) can be used.

Efficient implementation techniques for constraint solving have been
studied extensively, e.g., for ASP solvers~\cite{leone2006dlv,GebKKS12book}.

\myparag{Applications of constraint satisfaction}
The generality and power of constraints allow them to be used for all
applications described previously, but constraints are particularly
important for applications beyond those and that require combinatorial
search.  Common kinds of search problems include planning and scheduling,
resource allocation, games and puzzles, and well-known NP-complete problems
such as graph coloring, k-clique, set cover, Hamiltonian cycle, and SAT.

%
%
%

We describe three example applications, in the domains of decision
making\notes{planning}, resource allocation, and games and puzzles.
They all require substantial use of general constraints and efficient
constraint solvers exploiting backtracking, backjumping, and other
optimizations.

\begin{description}
\item \descheading{Enterprise decision making---Prescriptive analysis\notes{Planning}}
  Prescriptive analysis suggests decision options that lead to optimized
  future actions.
  It is an advanced component of enterprise software.
  For example, for planning purposes, LogicBlox supports prescriptive
  analysis\notes{planning} using the same Datalog-based language as for BI
  and OLTP~\cite{green12logicblox,aref2015design}.

  The data are objects and relations, same as used for BI, but may include,
  in particular, costs and other objective measures.
  Constraints capture restrictions among the objects and relations.  When
  all data values are provided, constraints
  can simply be checked. 
  When some data values are not provided, different choices for those
  values can be explored, and values that lead to certain maximum or
  minimum objective measures may be prescribed for deciding future actions.
  Efficient implementations can utilize the best constraint solvers based
  on the kinds of constraints used.
%
%

  LogicBlox's integrated solution to decision making\notes{planning} based
  on BI and OLTP
  has led to significant success.  For example, for a Fortune 50 retailer
  with over \$70 billion in revenue and 
  with products %
  available through over 2,000 stores and digital channels,
  the solution 
  processes 3 terabytes of data on daily, weekly, and monthly cycles,
  deciding exactly what products to sell in what stores in what time
  frames; this reduces a multi-year cycle of a challenging task for a large
  team of merchants and planners to an automatic 
  process and significantly increases profit
  margins~\cite{logicblox15plan}.

\item \descheading{Resource allocation---Workforce management (WFM) in Port
    of Gioia Tauro}
  WFM handles activities needed to maintain a productive workforce.  The
  WFM system for automobile logistics in the Port of Gioia Tauro, the
  largest transshipment terminal in the Mediterranean, allocates available
  personnel of the seaport such that cargo ships mooring in the port are
  properly handled~\cite{ricca2012team,LeoneRicca2015asp}.  It was
  developed using the DLV system~\cite{leone2006dlv}.
  
  The data include employees of different skills, cargo ships of different
  sizes and loads, teams and roles to be allocated, and many other objects
  to be constrained, e.g., workload of employees, heaviness of roles, and
  contract rules.  Constraints include matching of available and required
  skills, roles, hours, etc., fair distribution of workload, turnover of
  heavy or dangerous roles, and so on.  The constraints are expressed using
  rules with disjunction in the conclusion, general negation, and
  aggregates.  The DLV system uses backtracking and a suite of efficient
  implementation techniques.

  This WFM system was developed by Exeura s.r.l.\ and has been adopted by
  the company ICO BLG operating automobile logistics in the Port of Gioia
  Tauro~\cite{LeoneRicca2015asp}, handling every day several
  ships\notes{vessels} of different sizes\notes{capacities}
  that moor in the port~\cite{ricca2012team}.

\item \descheading{Games and puzzles---N-queens} 
  We use a small example in a large class of problems.
  The n-queens puzzle is the 
  problem of placing \c{n} queens on a chessboard of \c{n}-by-\c{n} squares
  so that no two queens threaten each other, i.e., no two queens share the
  same row, column, or diagonal.  The problem is old, well-studied, and can
  be computationally quite expensive~\cite{bell2009survey}.
  
  The allowed placements of queens can be specified as logic rules with
  constraints.  Naively enumerating all possible combination of positions
  and checking the constraints is prohibitively expensive\notes{ even for
    eight queens}.  More efficient solutions use backtracking, and
  furthermore backjumping, to avoid impossible placement of each next queen
  as soon as possible.
  Stronger forms of constraints may also be specified to help prune the
  search space further~\cite{GebKKS12book}.
  For example, backtracking can solve for one or two scores of queens in an
  hour, but backjumping and additional constraints help an ASP system like
  Clingo solve for 5000 queens in 3758.320 seconds of CPU
  time~\cite{schaub14slides}.

  Many other games and puzzles can be specified and solved in a similar
  fashion.  Examples are all kinds of crossword puzzles, Sudoku, Knight's
  tour, nonograms, magic squares, dominos, coin puzzles, graph
  coloring, 
  palindromes, among many others,
  e.g.,~\cite{demoen05first,edmunds15puzzles,hett15prolog,malita15logic,kje15lpicat}.
  %
  %
\end{description}

\section{Further extensions,  applications, and discussion}
\label{sec-ext}
%
%
%

We discuss additional language extensions and applications, summarize
applications based on the key abstractions used, touching on example logic
programming systems, and finally put the abstractions into the perspective
of programming paradigms.

\myparag{Extensions}
Many additional extensions to logic languages have been studied.  Most of
them can be viewed as abstractions that capture common patterns in classes
of applications, to allow applications to be expressed more easily.
Important 
extensions include: 
\begin{itemize}

\item regular-expression paths, a higher-level abstraction for
  commonly-used linear recursion;

\item updates, for real-world applications that must handle changes; 

\item time, for expressing changes over time, as an alternative to
  supporting updates directly; 

\item probability, to capture uncertainty in many challenging applications; and

\item higher-order logic, to support applications that require meta-level
  reasoning.

\end{itemize}
We discuss two of the most important extensions below:

\begin{description}

\item \descheading{Regular-expression paths}
  A regular-expression path relates two objects using regular expressions
  and extensions.  It allows repeated joins of a binary relation to be
  expressed more easily and clearly than using recursion; such joins
  capture reachability and are commonly used.  For example,
  \c{is\_ancestor(X,Y)}, defined in Section~\ref{sec-recursion} using two
  rules including a recursive rule, can now be defined simply as below; it
  indicates that there are one or more \c{is\_parent} relationships in a path
  from \c{X} to \c{Y}:
\begin{code}
    is_ancestor(X,Y) \IF is_parent+(X,Y).
\end{code}
  This is also higher-level than using recursion, because the recursive
  rule has to pick one of three possible forms below: with \c{is\_parent} 
  on the left, as seen before; with \c{is\_parent} on the right; 
  and with both conjuncts using \c{is\_ancestor}.
\begin{code}
    is_ancestor(X,Y) \IF is_parent(X,Z), is_ancestor(Z,Y).
    is_ancestor(X,Y) \IF is_ancestor(X,Z), is_parent(Z,Y).
    is_ancestor(X,Y) \IF is_ancestor(X,Z), is_ancestor(Z,Y).
\end{code}
  Depending on the data, the performance of these forms can be
  asymptotically different in most implementations.

  Regular-expression paths have many important applications including all
  those in Section~\ref{sec-recursion}, especially graph queries, with also
  parametric extensions for more general relations, not just binary
  relations~\cite{deM+03,Liu+04PRPQ-PLDI,LiuSto06GraphQL-PADL,Tek+10ImplGraphQL-PPDP}.

\item \descheading{Updates}
  An update, or action, can be expressed as a predicate that captures the
  update, e.g., by relating the values before and after the update and the
  change in value.
  The effect of the update could be taken immediately after the predicate
  is evaluated, similar to updates in common imperative languages, but this
  leads to lower-level control flows that are harder to reason about.
  Instead, it is better for the update to take effect as part of a
  transaction of multiple updates that together satisfy high-level logic
  constraints.
  For example, with this approach, the following rule means that
  \c{adopted\_by\_from} holds if the updates \c{add\_child} and
  \c{del\_child} and the check \c{adoption\_check} happen as a transaction.
\begin{code}
    adopted_by_from(C,X,Y) \IF add_parent(X,C), del_parent(Y,C), adoption_check(C,X,Y).
\end{code}
  It ensures at a high-level that certain bad things won't happen, e.g., 
  no child would end with one fewer parent or one more parent than
  expected.
  %
  Transaction logic is an extension of logic rules for reasoning about and
  executing transactional state changes~\cite{BonKif94}.
  LPS, a Logic-based approach to Production Systems, captures state changes
  by associating timestamps with facts and events, and this is shown to
  correspond to updating facts directly~\cite{kow15reactive}.

  Logic languages with updates have important applications in enterprise
  software~\cite{green12logicblox,aref2015design}.  
  Transaction logic can also help in planning~\cite{basseda2014planning}.

%
%
%

%
%
%
%

\end{description}

Additional implementation support can help enhance applications and enable
additional applications.  A particular helpful feature is to record
justification or provenance information during program
execution~\cite{roychoudhury2000justif,damasio2013justif}, providing
explanations for how a result was obtained.  The recorded information can
be queried to improve understanding and help debugging.
%
%

\myparag{Additional applications}
Many additional applications 
have been developed using logic programming, especially including
challenging applications that need recursion and those that furthermore
need constraint.

Table~\ref{tab-appl} lists example application areas with example
application problems organized based on the main abstractions used.
Note that application problems can often be reduced to each other,  
and many other problems can be reduced to the problems in the table.  For
example, model checking a property of a
system~\cite{Clarke+99,Clarke:Grumberg:Long:94} can be
reduced to planning, where the goal state is a state violating the property
specified, so a plan found by a planner corresponds to an error trace found
by a model checker~\cite{MCAP14,APMC14}.
Administrative policy analysis also has correspondences to planning, by
finding a sequence of actions to achieve the effect of a security
breach~\cite{stoller2011symbolic}.

\begin{table}[htb]
  \centering
\begin{tabular}{@{~}m{10.9ex}@{~}||@{~}m{21.5ex}@{~}|@{~}m{26.5ex}@{~}|@{~}m{24ex}@{~}}
Area
& Using join
& Using recursion
& Using constraint
\\\hline\hline
       
Data\newline{} management
& business intelligence,*\newline 
  many database\newline \Fsptwo join queries
& route queries,\newline 
  many database\newline \Fsptwo recursive queries
& data cleaning,\newline 
  data repair 
\\\hline

Knowledge management
& ontology\newline \Fsptwo management*
& ontology analysis 
& reasoning with\newline \Fsptwo knowledge 
\\\hline

Decision support
& 
& supply-chain management,\newline 
  market analysis 
& prescriptive analysis,*\newline
  planning, scheduling,\newline 
  resource allocation*
\\\hline

Linguistics 
& 
& 
  text processing,*\newline
  context-free parsing,\newline semantic analysis
& context-sensitive\newline \Fsptwo analysis,\newline 
  deep semantics analysis 
\\\hline



Program analysis
& type checking,\newline many local analyses
& pointer analysis,*\newline type analysis,\newline 
  many dependency analyses
& type inference
  ,\newline many constraint-based\newline \Fsptwo analyses
\\\hline


Security 
& role-based\newline \Fsptwo access control*
& trust management,*\newline 
  hierarchical role-based\newline \Fsptwo access control
& administrative policy\newline \Fsptwo analysis,\newline cryptanalysis
\\\hline

Games and puzzles
&
& Hanoi tower, \newline many recursion problems
& n-queens,*\newline Sudoku,\newline many constraint puzzles
\\\hline

Teaching
& course management
& course analysis 
& question analysis,\newline problem diagnosis,\newline test generation
\\\hline

\end{tabular}

\caption{Example application areas with example application problems
  organized based on the main abstractions used.
  Applications discussed in some detail in this article are marked 
  with an asterisk.}
\label{tab-appl}
\end{table}

Table~\ref{tab-appl} is only a small sample of the application areas, with
example application problems or kinds of application problems in those
areas.  Many more applications have been developed, in many more areas,
using systems that support variants of the abstractions with different
implementations.  Some examples are:
\begin{itemize}

\item XSB has also been used to develop applications for 
  immunization survey~\cite{burton12}, standardizing data, spend analysis,
  etc.~\cite{xsb15case},
  %
  and it is discussed in many publications\footnote{
    A Google Scholar search with \c{+XSB +''logic programming''} returns
    over 2300 results, July 2, 2017.}.

\item LogicBlox has also been used to create solutions for predicting
  consumer demand, optimizing supply chain, etc.~\cite{logicblox15sol} and
  more~\cite{green12logicblox}.

\item ASP systems have been used in bioinformatics, hardware design, music
  composition, robot control, tourism, 
  and many other application areas~\cite{wasp05,schaub11asp}, including
  part of a decision support system for the Space Shuttle flight
  controller~\cite{nogueira2001prolog,balduccini2005model}.
%
%
%

\item Logic systems have been developed 
  for additional applications, e.g.,
  PRISM~\cite{sato1997prism} and ProbLog~\cite{de2007problog} for
  probabilistic models; 
  %
  XMC~\cite{rama2000xmc} and ProB~\cite{leuschel2008prob}
  for verification;
  %
  and NDlog~\cite{loo09decl}, Meld~\cite{ashley-rollman-iclp09},
  Overlog~\cite{alvaro2010declare}, and Bloom~\cite{bloom} for network and
  distributed algorithms.



\end{itemize}
Languages and systems with more powerful features such as constraints for
general applications are often also used in less challenging application
areas such as those that need only join queries.  For example, DLV has also
been used in ontology management~\cite{ricca2009ontodlv}.

\myparag{Additional discussion on abstractions}
We give an overview of the main abstractions in the larger picture of
programming paradigms, to help put the kinds of applications supported into
broader perspective.

The three main abstractions---join, recursion, and constraint---correspond
generally to more declarative programming paradigms.  Each is best known in
its corresponding main programming community:
\begin{itemize}

\item Join in database programming.  Database systems have join at the core
  but support restricted recursion and constraints in practice.

\item Recursion in functional programming.  Functional languages have
  recursion at the core but do not support high-level join or constraints.

\item Constraint in logic programming.  Logic engines support both join and
  recursion at the core, and have increasingly supported constraints at the
  core as well.

\end{itemize}
The additional extensions help further raise the level of abstraction and
broaden the programming paradigms supported:
\begin{itemize}

\item Regular-expression paths raise the level of abstraction over
  lower-level linear recursion.

\item Updates, or actions, are the core of imperative programming; they
  help capture real-world operations even when not used in low-level
  algorithmic steps.

\item Time, probability, higher-order logic, and many other features
  correspond to additional arguments, attributes, or abstractions about
  objects and relationships.


\end{itemize}
One main paradigm not yet discussed is object-oriented programming.
Orthogonal to data and control abstractions, objects in common
object-oriented languages provide a kind of module abstraction,
encapsulating both data structures and control structures in objects and
classes.  Similar abstractions have indeed been added to logic languages as
well.  For example, F-logic extends traditional logic programming with
objects~\cite{KifLW95} and is supported in Flora-2~\cite{flora14}; it was
also the basis of a highly scalable commercial system,
Ontobroker~\cite{ontobroker}, and a recent industry suite,
Ergo~\cite{Grosof15ruleml}.  For another example, ASP has been extended
with object constructs in OntoDLV~\cite{ricca2009ontodlv}.

Finally, building practical applications requires powerful libraries and
interfaces for many standard functionalities.  Many logic programming
systems provide various such libraries.  For example, SWI-Prolog has
libraries for constraint logic programming, multithreading, interface to
databases, GUI, a web server, etc, as well as development tools and
extensive documentation.

%
%
%
%
%
%

\section{Related literature and future work}
\label{sec-relate}
There are many overview books and articles about logic programming in
general and applications of logic programming in particular.  
This article differs from prior works by studying the key abstractions and
their implementations as the driving force underlying vastly different
application problems and application areas.

%

Kowalski~\cite{kow14lp} provides an extensive overview of the development
of logic programming.  It describes the historical root of logic
programming,
starting from resolution theorem-proving; the procedural interpretation and
semantics of rules with no negated hypotheses, called Horn clause programs;
negation as failure, including completion semantics, stratification,
well-founded semantics, stable model semantics, and ASP; as well as logic
programming involving abduction, constraints, and argumentation.
It focuses on three important issues: logic programming as theorem proving
vs.\ model generation, with declarative vs.\ procedural semantics, and
using top-down vs.\ bottom-up computation.
%
Our description of abstractions and implementations aims to separate
declarative semantics from procedural implementations.

Other overviews and surveys about logic programming in general
include some that cover a collection of topics together
and some that survey different topics separately.
%
Example collections discuss the first 25 years of logic programming from
1974~\cite{apt1999lp25} and the first 25 years of the Italian Association
of Logic Programming from 1985~\cite{dovier2010lp25}.
Example topics surveyed separately include %
logic programming semantics~\cite{fitting2002fixpoint}, %
complexity and expressive power~\cite{dantsin2001complexity}, %
constraints~\cite{jaf+1994constr}, %
ASP and DLV~\cite{grasso2013asp}, %
deductive
databases~\cite{Ceri:Gottlob:Tanca:90,AbiHulVia95,RamUll95survey,Min+14history}, %
and many more.
Our description of abstractions and implementations is only a highly
distilled overview of the core topics.
%

Overviews and surveys about logic programming applications in particular
are spread across many forums.
Example survey articles include %
an early article on Prolog applications~\cite{roth1993practical}, %
%
%
DLV applications~\cite{grasso2009some,grasso2011asp,LeoneRicca2015asp},
applications in Italy~\cite{dal2010lp25}, 
emerging applications~\cite{huang2011datalog}, 
%
and a dedicated workshop AppLP---Applications of Logic
Programming~\cite{WarLiu17AppLP-arxiv}.
For example, the early article~\cite{roth1993practical} describes six
striking practical applications of Prolog that replaced and drastically
improved over systems written previously using Fortran, C++, and Lisp.
Example collections of applications on the Web include %
one at TU Wien~\cite{wasp05}, 
one by Schaub~\cite{schaub11asp}, 
and some of the problems in various competitions, e.g., as described
by Gebser et al.~\cite{gebser17sixth}.
%
%
%
%
%
%
%
%
We try to view the applications by the abstractions and implementations
used, so as to not be distracted by specific details of very different
applications.\notes{; we also give an overview together with the
  application areas afterwards.}

There are also many articles on specific applications or specific classes
of applications.
%
Examples of the former include team building~\cite{ricca2012team}, program
pointer analysis~\cite{smara15pointer}, and others discussed in this
article.
Examples of the latter include applications in software
engineering~\cite{cian1995report},
DLV applications in knowledge management~\cite{grasso2009some}, %
and IDP applications in data mining and machine
learning~\cite{bruynooghe2014predicate}.
We used a number of such specific applications as examples and described
some of them in slightly more detail to illustrate the common technical
core in addition to the applications per se.

\paragraph{Directions for future work.}
There are several main areas for future study:
(1) more high-level abstractions that are completely declarative,
(2) more efficient implementations with complexity guarantees, and
(3) more unified and standardized languages and frameworks with rich libraries.
These will help many more 
applications to be created in increasingly complex problem domains.

\subsection*{Acknowledgment}

I would like to thank\notes{changed by Kifer: This article is dedicated to}
David S.\ Warren\notes{,} for his encouragement over the years at Stony
Brook, and his patient and stimulating explanations about logic programming
in general and XSB implementation in particular.
I am grateful to Molham Aref, Francesco Ricca, and David Warren for helpful
suggestions and additional information about applications using LogicBlox,
DLV, and XSB, respectively.
I thank Molham Aref and others at LogicBlox, Jon Brandvein, Christopher
Kane, Michael Kifer, Bob Kowalski, Bo Lin, Francesco Ricca, Scott Stoller,
Tuncay Tekle, David Warren, Neng-Fa Zhou, and anonymous reviewers for
helpful comments on drafts of this article.

\def\bibdir{../../../bib}                    
{\renewcommand{\baselinestretch}{-.7}\footnotesize
\bibliography{\bibdir/strings,\bibdir/liu,\bibdir/IC,\bibdir/PT,\bibdir/PA,\bibdir/Veri,\bibdir/Lang,\bibdir/Algo,\bibdir/Perform,\bibdir/DB,\bibdir/SE,\bibdir/Sys,\bibdir/Sec,\bibdir/misc,\bibdir/crossref}
\bibliographystyle{alpha}
}
\end{document}